\documentclass[twocolumn, twocolappendix]{aastex7}

\usepackage{amsmath}
\usepackage{xcolor}
\usepackage{comment}
\usepackage{graphicx}
\usepackage{adjustbox}
\usepackage{hyperref}
\usepackage{array}
\newcolumntype{s}{@{\hskip 1pt}c@{\hskip 1pt}} 
\newsavebox\ltmcbox

\newcommand{\PSUAA}{Department of Astronomy \& Astrophysics, 525 Davey Laboratory, 251 Pollock Road, Penn State, University Park, PA, 16802, USA}
\newcommand{\PSUCEHW}{Center for Exoplanets and Habitable Worlds, 525 Davey Laboratory, 251 Pollock Road, Penn State, University Park, PA, 16802, USA}
\newcommand{\PSUARC}{Astrobiology Research Center, 525 Davey Laboratory, 251 Pollock Road, Penn State, University Park, PA, 16802, USA}
\newcommand{\PSETI}{Penn State Extraterrestrial Intelligence Center, 525 Davey Laboratory, 251 Pollock Road, Penn State, University Park, PA, 16802, USA}
\newcommand{\PSUICDS}{Institute for Computational and Data Sciences, Penn State, University Park, PA, 16802, USA}
\newcommand{\PSUCASt}{Center for Astrostatistics, 525 Davey Laboratory, 251 Pollock Road, Penn State, University Park, PA, 16802, USA}
\newcommand{\UA}{Steward Observatory, University of Arizona, 933 N.\ Cherry Ave, Tucson, AZ 85721, USA}

\newcommand{\Penn}{Department of Physics and Astronomy, University of Pennsylvania, 209 S 33rd St, Philadelphia, PA 19104, USA}

\newcommand{\MacquarieAAO}{Australian Astronomical Optics, Macquarie University, Balaclava Road, Sydney, NSW 2109, Australia.}
\newcommand{\Macquarie}{School of Mathematical and Physical Sciences, Macquarie University, Balaclava Road, North Ryde, NSW 2109, Australia}
\newcommand{\MacquarieCentre}{Astrophysics and Space Technologies Research Centre, Macquarie University, Balaclava Road, North Ryde, NSW 2109, Australia}

\newcommand{\JPL}{Jet Propulsion Laboratory, California Institute of Technology, 4800 Oak Grove Drive, Pasadena, California 91109}

\newcommand{\UCI}{Department of Physics \& Astronomy, The University of California, Irvine, Irvine, CA 92697, USA}
\newcommand{\Carleton}{Carleton College, One North College St., Northfield, MN 55057, USA}
\newcommand{\Carnegie}{Carnegie Science Earth and Planets Laboratory, 5241 Broad Branch Road, NW, Washington, DC 20015, USA}

\newcommand{\FlatironCCA}{Center for Computational Astrophysics, Flatiron Institute, 162 Fifth Avenue, New York, NY 10010, USA}

\newcommand{\TIFR}{Department of Astronomy and Astrophysics, Tata Institute of Fundamental Research, Homi Bhabha Road, Colaba, Mumbai 400005, India}

\newcommand{\UAm}{Anton Pannekoek Institute for Astronomy, 904 Science Park, University of Amsterdam, Amsterdam, 1098 XH}
\newcommand{\UIUC}{Department of Astronomy, University of Illinois at Urbana-Champaign, Urbana, IL 61801, USA}
\newcommand{\Amherst}{Department of Physics and Astronomy, Amherst College, 25 East Drive, Amherst, MA 01002, USA}
\newcommand{\PSUICS}{Institute for Computational and Data Sciences, Penn State, University Park, PA, 16802, USA}
\newcommand{\NOAO}{U.S. National Science Foundation National Optical-Infrared Astronomy Research Laboratory, 950 N.\ Cherry Ave., Tucson, AZ 85719, USA}

\submitjournal{the Astronomical Journal: Accepted February 20, 2026}

\shorttitle{HD 190360 \text{$\mathrm{d}$}}
\shortauthors{Giovinazzi et al.}

\begin{document}

\title{The NEID Earth Twin Survey. IV. \\ Confirming an 89~d, $m\sin i=10~\mathrm{M_\oplus}$ Planet Orbiting a Nearby Sun-like Star}

\author[0000-0002-0078-5288]{Mark R. Giovinazzi}
\affiliation{\Amherst}
\email{mgiovinazzi@amherst.edu}

\author[0000-0003-0199-9699]{Evan Fitzmaurice}
\affiliation{\PSUAA}
\affiliation{\PSUCEHW}
\affiliation{\PSUICS}
\email{exf5296@psu.edu}

\author[0000-0002-5463-9980]{Arvind F.\ Gupta}
\affiliation{\NOAO}
\email{arvind.gupta@noirlab.edu}

\author[0000-0003-0149-9678]{Paul Robertson}
\affiliation{\UCI}
\email{paul.robertson@uci.edu}

\author[0000-0001-9596-7983]{Suvrath Mahadevan}
\affiliation{\PSUAA}
\affiliation{\PSUCEHW}
\affiliation{\PSUARC}   
\email{suvrath@astro.psu.edu}

\author[0000-0001-6545-639X]{Eric B.\ Ford}
\affiliation{\PSUAA}
\affiliation{\PSUCEHW}
\affiliation{\PSUICDS}
\affiliation{\PSUCASt}
\email{ebf11@psu.edu}

\author[0000-0003-0353-9741]{Jaime A. Alvarado-Montes}
\affiliation{\MacquarieAAO}
\affiliation{\MacquarieCentre}
\email{jaime.alvaradomontes@mq.edu.au}

\author[0000-0003-4384-7220]{Chad F.\ Bender}
\affiliation{\UA}
\email{cbender@arizona.edu}

\author[0000-0002-6096-1749]{Cullen H.\ Blake}
\affiliation{\Penn}
\email{chblake@sas.upenn.edu}

\author[0000-0002-3610-6953]{Jiayin Dong}
\affiliation{\FlatironCCA}
\affiliation{\UIUC}
\email{jdong@flatironinstitute.org}

\author[0000-0002-3853-7327]{Rachel B. Fernandes}
\affiliation{\PSUAA}
\affiliation{\PSUCEHW}
\altaffiliation{President's Postdoctoral Fellow}
\email{rbf5378@psu.edu}

\author[0000-0003-1312-9391]{Samuel Halverson}
\affiliation{\JPL}
\email{samuel.halverson@jpl.nasa.gov}

\author[0000-0002-7127-7643]{Te Han}
\affiliation{\UCI}
\email{teh2@uci.edu}

\author[0000-0001-8401-4300]{Shubham Kanodia}
\affiliation{\Carnegie}
\email{skanodia@carnegiescience.edu}

\author[0000-0001-9626-0613]{Daniel M.\ Krolikowski}
\affiliation{\UA}
\email{krolikowski@arizona.edu}

\author[0000-0002-9632-9382]{Sarah E.\ Logsdon}
\affiliation{\NOAO}
\email{sarah.logsdon@noirlab.edu}

\author[0000-0001-8720-5612]{Joe P.\ Ninan}
\affiliation{\TIFR}
\email{indiajoe@gmail.com}

\author[0000-0001-8127-5775]{Arpita Roy}
\affiliation{Astrophysics \& Space Institute, Schmidt Sciences, New York, NY 10011, USA}
\email{arpita308@gmail.com}

\author[0000-0002-4046-987X]{Christian Schwab}
\affiliation{\Macquarie}
\email{mail.chris.schwab@gmail.com}

\author[0000-0001-7409-5688]{Gudmundur Stefansson}
\affiliation{\UAm}
\email{g.k.stefansson@uva.nl}

\author[0000-0002-4788-8858]{Ryan C. Terrien}
\affiliation{\Carleton}
\email{rterrien@carleton.edu}

\author[0000-0001-6160-5888]{Jason T.\ Wright}
\affiliation{\PSUAA}
\affiliation{\PSUCEHW}
\affiliation{\PSETI}
\email{astrowright@gmail.com}

\correspondingauthor{Mark R. Giovinazzi}
\email{mgiovinazzi@amherst.edu}

\begin{abstract}

We present the confirmation of HD 190360~d, a warm ($P=88.690^{+0.051}_{-0.049}~\mathrm{d}$), low-mass ($m\sin i=10.23^{+0.81}_{-0.80}~\mathrm{M_\oplus}$) planet orbiting the nearby ($d=16.0$~pc), Sun-like (G7) star HD 190360. We detect HD 190360~d at high statistical significance even though its radial velocity (RV) semi-amplitude is only $K=1.48\pm0.11~\mathrm{m~s^{-1}}$. Such low-amplitude signals are often challenging to confirm due to potential confusion with low-amplitude stellar signals. The HD 190360 system previously had two known planets: the $1.7~\mathrm{M_J}$ (true mass) HD 190360~b on a $7.9$~yr orbit and the $21~\mathrm{M_\oplus}$ (minimum mass) HD 190360~c on a $17.1$~d orbit. Here, we present an in-depth analysis of the HD 190360 planetary system that comprises more than 30 years of RV measurements and absolute astrometry from the Hipparcos and Gaia spacecrafts. Our analysis uses more than 1400 RVs, including nearly 100 from NEID. The proper motion anomaly as measured by these two astrometric missions solves for the dynamical mass of HD 190360~b and contributes to our understanding of the overall system architecture, while the long baseline of RVs enables the robust characterization of HD 190360~c and confirms the discovery of HD~190360~d.

\end{abstract}


\section{Introduction} \label{sec:intro}

\setcounter{footnote}{0}

Since the discovery of 51 Pegasi~b \citep{1995Natur.378..355M}, the radial velocity (RV) method for detecting and characterizing exoplanets has directly contributed to the confirmation of more than 2300 exoplanets\footnote{\href{https://exoplanetarchive.ipac.caltech.edu}{https://exoplanetarchive.ipac.caltech.edu}}. More than 50\% of these have periods $<15~\mathrm{d}$, while fewer than 10\% have periods $>5~\mathrm{yr}$. Moreover, $\lesssim10\%$ of RV-discovered planets have RV semi-amplitudes ($K$) smaller than $2~\mathrm{m~s^{-1}}$. However, long baselines of RV observations, paired with increasingly precise Doppler spectrometers such as NEID \citep{NEID_design}, KPF \citep{2016SPIE.9908E..70G}, EXPRES \citep{2016SPIE.9908E..6TJ}, MAROON-X \citep{2018SPIE10702E..6DS}, CARMENES \citep{2018SPIE10702E..0WQ}, and ESPRESSO \citep{2021A&A...645A..96P}, are beginning to enable the detection of planets with low-RV amplitudes owed to their lower masses and/or longer periods \citep[e.g.,][]{2022AJ....163..218H, 2024A&A...684A.117D, HD86728}. 

The NEID Earth Twin Survey (hereafter NETS; \citealt{NETS, NETSII, NETSIII}) is a Guaranteed Time Observations (GTO) program for the NEID spectrograph with the goal of searching for low-mass exoplanets. Among the 41 NETS targets is HD 190360, a nearby ($d=16.0~\mathrm{pc}$) G-type star. HD 190360 hosts two previously-discovered planets -- a cold Jupiter HD 190360~b \citep{HD190360_elodie_afoe} and a warm Neptune HD 190360~c \citep{2005ApJ...632..638V} -- and has more recently been investigated for evidence of a third companion \citep[e.g.,][]{2017PhDT.......136F, Hirsch_AO, 2021ApJS..255....8R}.

\cite{2017PhDT.......136F} first reported a low-amplitude periodic signal of $P=88.69\pm0.13~\mathrm{d}$ in the residuals of a two-planet fit to HD 190360 using more than 20 years of RVs. \cite{Hirsch_AO} later identified the same signal ($P=88.9^{+0.2}_{-0.1}~\mathrm{d}$) as a planet candidate, noting it as suspiciously close to the 1/4 annual harmonic. \cite{2021ApJS..255....8R} independently determined $P=90.34^{+0.11}_{-0.13}~\mathrm{d}$ but referred to it as an annual and/or instrumental systematic, likely as a result of its proximity to the aforementioned 1/4 annual harmonic. A residual period in this vicinity ($P=88.662^{+0.065}_{-0.064}~\mathrm{d}$) was also recovered by \cite{Harada_2025}. Most recently, \cite{NETSII} used new high-precision NEID RVs to highlight the same periodic signal at $P=88.818\pm0.069~\mathrm{d}$ with an amplitude consistent with $m\sin i\approx10~\mathrm{M_\oplus}$, favoring the case for a new low-mass planet, hereafter referred to as HD 190360~d, over an annual or systematic effect as the source of the signal. In this work, we revisit the same signal with additional NEID data and careful modeling to confirm its interpretation as a planetary companion.

Short-period ($\lesssim100~\mathrm{d}$) super-Earths and sub-Neptunes ($m\lesssim20~\mathrm{M_\oplus}$) are among the most common class of planet found so far (e.g., \citealt{2021JGRE..12606639B} and references therein). Evidence suggests that inner low-mass planets, like HD 190360~d, are more common in systems with cold Jupiters, implying that giant outer planets may be a key ingredient in fostering long-term stability within the inner disk \citep[e.g.,][]{Zhu__2018, 2019AJ....157...52B}. This is particularly true for stars of higher metallicity \citep[e.g.,][]{Weiss_2023, bryan2024friendsfoesstrongcorrelation, zhu2024metallicitydimensionsuperearthcold}, such as HD 190360 ($\mathrm{\left[Fe/H\right]}=0.26\pm0.06$~dex; \citealt{Hirsch_AO}).

Our paper is organized in the following way. In Section \ref{sec:data}, we describe the complete dataset used to perform our analysis. Section \ref{sec:star} examines the known stellar properties of HD 190360 and assesses the impact of its wide binary component. We outline the framework of our global Markov Chain Monte Carlo (MCMC) analysis in Section \ref{sec:analysis} and present the results in Section \ref{sec:results}. Both the fidelity and impact of our confirmation of HD 190360~d are discussed in Section \ref{sec:discussion}. We summarize our findings in Section \ref{sec:conclusion}.

\section{Data} \label{sec:data}

Our dataset consists of RVs from multiple spectrographs and absolute astrometry from the Hipparcos-Gaia Catalog of Accelerations (HGCA; \citealt{hgca_dr2, hgca_edr3}). Here, the long-term proper motion anomaly observed between the epochs of Hipparcos and Gaia is used to constrain the dynamical mass and orbital architecture of the system's outer planet, HD 190360~b \citep{NETSII}. We obtain archival RV measurements from ELODIE, AFOE, HIRES, Hamilton, and APF. In addition, we publish new NEID observations. We also consider publicly available TESS \citep{TESS} photometry of HD 190360.

\subsection{TESS} \label{subsec:tess}

HD 190360 was observed by TESS in Sectors 54, 55, 74, 75, 81, and 82. We performed a Lomb-Scargle \citep{lomb, scargle} periodogram and a transit least squares \citep{2019A&A...623A..39H} search for periodic variability and transit-like signals over periods of 1-30~d and find no compelling evidence for either.

\subsection{NEID} \label{subsec:neid}

NEID is a high-resolution ($R>110,000$), ultra-stable, fiber-fed echelle spectrograph with wavelength coverage greater than $\lambda=380-930~\mathrm{nm}$ and internal RV precision better than $\sim27~\mathrm{cm~s^{-1}}$ \citep{NEID_error_budget}. As one of the NETS target stars, HD 190360 has been monitored with NEID since 2021. In total, we use 93 NEID exposures from 2021 May 31 - 2024 Nov 18. This is 54 more observations than those published in \cite{NETSII}. All of these RVs were extracted using version 1.3 of the NEID Data Reduction Pipeline\footnote{\url{ https://neid.ipac.caltech.edu/docs/NEID-DRP/}} (DRP). In our analysis, we also consider the activity indicators measured from each NEID spectrum (e.g., Ca II H and K, NaI, and H$\alpha$; see NEID DRP documentation for full list of indicators).

Our NEID data are split into three RV eras (Run 0.5, Run 1, and Run 2) following the RV zero-point offsets described in \cite{NETSIII}. Run 0.5 corresponds to data taken prior to 2021 August 7 before a distinct change in the average line profile was observed (see \citealt{NETSIII} for a more detailed description). The break between Run 1 and Run 2 is the result of the Contreras fire atop Kitt Peak that ceased NETS observations at the WIYN 3.5~m telescope\footnote{The WIYN Observatory is a joint facility of the NSF’s National Optical-Infrared Astronomy Research Laboratory, Indiana University, the University of Wisconsin-Madison, Pennsylvania State University, and Princeton University.} between UT 2022 June 14 and 2022 November 30. The NEID RVs used in this work are made up of seven points from Run 0.5, 22 points from Run 1, and 64 points from Run 2. We present a subset of the NEID RVs in Table \ref{tab:neid_rvs}, and include all NEID data used here in the form of a machine readable table.

\begin{deluxetable*}{rccc}
\tablecaption{NEID RVs for HD~190360\label{tab:neid_rvs}}
\tablehead{
\colhead{~~BJD$_{\mathrm{TDB}}~\left[\mathrm{d}\right]$~~} &
\colhead{~~RV~$\left[\mathrm{m~s^{-1}}\right]$~~} &
\colhead{~~$\sigma_{\mathrm{RV}}$~$\left[\mathrm{m~s^{-1}}\right]$~~} &
\colhead{~~Run \#~~}
}
\startdata
2459365.7961 & -45281.55 & 0.38 & 0.5 \\
2459365.9056 & -45281.81 & 0.21 & 0.5 \\
2459370.9510 & -45271.63 & 0.54 & 0.5 \\
2459370.9646 & -45273.16 & 0.21 & 0.5 \\
2459383.7366 & -45284.16 & 0.21 & 0.5 \\
2459383.8119 & -45283.96 & 0.21 & 0.5 \\
2459424.8292 & -45280.76 & 0.20 & 0.5 \\
2459474.7598 & -45279.57 & 0.32 & 1 \\
2459478.8295 & -45286.71 & 0.46 & 1 \\
2459479.6803 & -45286.55 & 0.32 & 1 \\
2459493.6561 & -45282.14 & 0.32 & 1 \\
2459497.6471 & -45288.87 & 0.32 & 1 \\
...~~~~~~~~~ & ... & ... & ... \\
2460583.7478 & -45312.89 & 0.30 & 2 \\
2460589.6409 & -45314.07 & 0.29 & 2 \\
2460591.6352 & -45316.02 & 0.30 & 2 \\
2460598.6864 & -45321.23 & 0.30 & 2 \\
2460601.7268 & -45311.15 & 0.32 & 2 \\
2460606.6454 & -45315.84 & 0.30 & 2 \\
2460608.6677 & -45317.52 & 0.30 & 2 \\
2460611.6182 & -45321.58 & 0.30 & 2 \\
2460619.6595 & -45314.24 & 0.30 & 2 \\
2460623.6412 & -45318.07 & 0.30 & 2 \\
2460628.6006 & -45322.80 & 0.30 & 2 \\
2460632.6344 & -45323.52 & 0.30 & 2 \\
\enddata
\end{deluxetable*}

\subsection{Archival RVs}

Our analysis includes 1312 previously-published RV measurements from a long history of observations. These measurements include 56 RVs from the ELODIE \citep{ELODIE} spectrograph and 13 RVs from the Advanced Fiber Optic Echelle (AFOE; \citealt{AFOE}) spectrograph, as well as the California Legacy Survey \citep{2021ApJS..255....8R}, an aggregate of dedicated RV searches for exoplanets around nearby FGKM stars which comprises Hamilton \citep{Hamilton}, the high-resolution echelle spectrometer (HIRES; \citealt{1994SPIE.2198..362V}), and the Automated Planet Finder (APF; \citealt{2014SPIE.9145E..2BR}). We use 146 RVs from Hamilton, 343 from HIRES (91 and 252 from before and after a 2004 CCD upgrade, respectively; \citealt{HIRES_upgrade}), and 754 from APF. The two HIRES datasets, like those from NEID described in Section \ref{subsec:neid}, are ultimately fit for as separate instruments to allow for a floating RV zero-point offset to be determined.

\section{Stellar Parameters} \label{sec:star}

HD 190360 is bright ($V=5.71$; \citealt{1991adc..rept.....G}) and among the closest ($d=16.0~\mathrm{pc}$; \citealt{gaia_dr3}) G-type stars (G7IV-V; \citealt{2006AJ....132..161G}) to our Sun. Further, it is one of fewer than 10 G stars within 16~pc known to host multiple planets. We adopt our stellar parameters from \cite{Hirsch_AO}, which used \texttt{SpecMatch-syn} \citep{specmatch_syn} and \texttt{isoclassify} \citep{isoclassify} to determine that HD 190360 is metal-rich ($\mathrm{\left[Fe/H\right]}=0.26\pm0.06$) with a mass ($M_*=0.99\pm0.04~\mathrm{M_\odot}$), temperature ($T_\mathrm{eff}=5560\pm100~\mathrm{K}$), and surface gravity ($\mathrm{log}g~\mathrm{\left[cgs\right]}=4.1\pm0.1$) near to that of our Sun. These values corroborate spectroscopically-derived ones from \cite{2016ApJS..225...32B}, which found $\mathrm{\left[Fe/H\right]}=0.19\pm0.01$, $M_*=0.92\pm0.12~\mathrm{M_\odot}$, $T_\mathrm{eff}=5549\pm25~\mathrm{K}$, and $\mathrm{log}g~\mathrm{\left[cgs\right]}=4.290\pm0.028$, as well as an age of $8.2\pm1.4~\mathrm{Gyr}$. \cite{1996ApJ...457L..99B} and \cite{2004ApJS..152..261W} reported estimated rotation periods of 38~d and 40~d for HD 190360, respectively, using an empirical $\mathrm{log}R'_\mathrm{HK}$ relationship derived in \cite{1984ApJ...279..763N}. \cite{Hojjatpanah_2019} determined HD 190360 to have a projected rotational velocity of $v\sin i=2.39\pm0.82~\mathrm{km~s^{-1}}$.

The primary star is in a wide ($\approx180''$, or nearly 3000~au) orbit with an M4.5 V \citep{2015A&A...577A.128A} companion, HD 190360 B (LHS 3509). Using available photometry in the $G$ \citep{2021A&A...649A...1G}, $V$ \citep{1991adc..rept.....G}, and $R$ \citep{2003AJ....125..984M, 2013AJ....145...44Z} bands, which most closely overlap the wavelength coverage of the RV instruments considered here, we estimate flux contrast ratios of order $10^{-3}-10^{-4}$. Given both its wide separation and small flux ratio, we conclude that the spectral contamination of HD 190360 B is negligible. To quantify the gravitational impact of HD 190360 B on our analysis, we use \texttt{binary\_mc}\footnote{\url{https://github.com/markgiovinazzi/binary_mc}} \citep{mark_giovinazzi_2025_15352084}, which employs a Monte Carlo approach to simulate 10 million random orbits of the HD 190360 AB binary system consistent with the stars' positions from Gaia. Assuming masses of $M_\mathrm{A}=0.99\pm0.04~\mathrm{M_\odot}$ as described above and $M_\mathrm{B}=0.20\pm0.02~\mathrm{M_\odot}$ derived from the Gaia-based photometric mass relationship presented in \citealt{2022AJ....164..164G}, we find the two stars to have a most likely period between $85,000-90,000~\mathrm{yr}$, with more than 99\% of orbits inducing an RV slope $<0.04~\mathrm{m~s^{-1}~yr^{-1}}$ and a most likely RV slope $\lesssim0.0035~\mathrm{m~s^{-1}~yr^{-1}}$. Given the negligible dynamical influence of the wide companion, we exclude it from our analysis outlined in Section \ref{sec:analysis}.


\section{MCMC Analysis} \label{sec:analysis}

To supervise our three-planet fit, we use \texttt{orvara} \citep{orvara}, which employs a parallel-tempering MCMC sampler that uses \texttt{ptemcee} \citep{emcee, ptemcee}. \texttt{orvara} jointly analyzes RVs, relative astrometry between resolved gravitationally-bound bodies, and absolute astrometry as estimated from three proper motion measurements: the proper motions as reported by each of Hipparcos \citep{Hipparcos_catalog} and Gaia \citep{Gaia_mission}, as well as a long-term proper motion constructed from the positions reported by each respective mission. We do not include HD 190360 B in our fit, and therefore do not utilize any relative astrometry.

We repeat the fitting routine for HD 190360 outlined in Section 4 of \cite{NETSII} and impose a Gaussian prior on the parallax based on the measurement published in the Gaia DR3 catalog with its error inflated by a factor of 1.28 following \cite{hgca_edr3}. We also adopt a Gaussian prior on the stellar mass based on the result from \cite{Hirsch_AO} described in Section \ref{sec:star}. The complete set of priors enforced in our MCMC analysis is provided in Table \ref{tab:priors}. Our MCMC is broadly initialized with 30 temperatures, 100 walkers, and 1,500,000 steps. We keep every 50th step and then remove the first 15,000 steps from each walker as burn-in, ensuring all chains are well-mixed. We assessed chain convergence using the Gelman-Rubin (GR) diagnostic \citep{GR_stat} and find that all fitted parameters have GR$<1.005$, indicating no evidence for lack of convergence. We also computed the integrated autocorrelation time, $\tau$, for each parameter using \texttt{emcee} and find $\tau\simeq 1$ for all free parameters, suggesting that the thinned chains are nearly independent.

\begin{deluxetable}{cc}
\caption{Priors used in the HD 190360 MCMC analysis}
\label{tab:priors}
\tablehead{\colhead{Parameter} & \colhead{Prior}}
\startdata
Parallax [mas]                    & $\mathcal{N}\left(62.4865,~0.0453\right)$ \\
Primary Mass [$\mathrm{M_\odot}$]          & $\mathcal{N}\left(0.99,~0.04\right)$ \\
Planet Mass [$\mathrm{M_\odot}$]        & $\mathrm{log_\mathrm{10}}~\mathcal{U}\left[10^{-6},~10^3\right]$ \\
Semimajor Axis [au]              & $\mathrm{log_\mathrm{10}}~\mathcal{U}\left[10^{-5},~2\times10^5\right]$ \\
RV Jitter [$\mathrm{m~s^{-1}}$]   & $\mathrm{log_\mathrm{10}}~\mathcal{U}\left[10^{-5},~20\right]$ \\
$\cos\left(\mathrm{Inclination}\left[\mathrm{rad}\right] \right)$           & $\mathcal{U}\left(0,~1\right)$ \\
$\sqrt{e}\sin\omega$            & $\mathcal{U}\left(-1,~1\right)$ \\
$\sqrt{e}\cos\omega$            & $\mathcal{U}\left(-1,~1\right)$ \\
Longitude of Ascending Node [rad] & $\mathcal{U}\left(-\pi,~3\pi\right)$ \\
Argument of Periastron [rad] & $\mathcal{U}\left(-\pi,~3\pi\right)$\\
RV Zero-point [$\mathrm{m~s^{-1}}$] & $\mathcal{U}\left(-\infty,~\infty\right)$ \\
\enddata
\tablecomments{Gaussian priors are used for parallax and host star mass. Log-flat (uniform in log-space) priors are used for all planet masses, semimajor axes, and instrumental RV jitters. Uniform priors are used for all other fitted parameters.}
\end{deluxetable}

To assess whether the three-planet model is statistically preferred over the simpler two-planet model reported in \cite{NETSII}, we fit both two- and three-planet models using our expanded RV dataset to enable a direct comparison. We then evaluate the models using the Bayesian Information Criterion (BIC) as a model selection metric, where

\begin{equation*}
    \mathrm{BIC}=k\log N - 2\log\mathcal{\hat{L}}.
\end{equation*}

Here, $k$ is the number of free parameters, $N$ is the number of data points, and $\log\mathcal{\hat{L}}$ is the maximum log-likelihood. Lower BIC values indicate preferred models, with the penalty term $k\log N$ disfavoring unnecessary complexity. For our dataset ($N=1411$, 1405 RVs plus six proper motion terms), the two-planet model has $k=24$ free parameters and $\mathrm{BIC=5246.42}$, while the three-planet model has $k=31$ free parameters and $\mathrm{BIC=5140.77}$. This yields a $\Delta\mathrm{BIC}=105.65$ in favor of the three-planet model, providing strong statistical support for the presence of HD 190360~d.

\section{Results} \label{sec:results}

Our joint analysis comprised of absolute astrometry and RVs not only refines the orbits of the two previously known planets, HD 190360~b and HD 190360~c, but also detects HD 190360~d with high fidelity. The full set of posterior summaries, including the median values, uncertainties based on the 16th and 84th percentiles of the marginalized posterior distributions, and the maximum a posteriori values, is provided in Table \ref{tab:planet_posteriors}.


\begin{deluxetable*}{ll|cc|cc|cc}
\label{tab:planet_posteriors}
\tablecaption{Median, 16th and 84th percentile credible intervals (Post.), and maximum a posteriori (MAP) values for the HD 190360 system.}

\tablehead{
\colhead{Parameter} & \colhead{Unit} &
\multicolumn{2}{c}{\textbf{HD 190360}} &
\multicolumn{2}{c}{} &
\multicolumn{2}{c}{} \\
\colhead{} & \colhead{} &
\colhead{Post.} & \colhead{MAP} &
\colhead{} & \colhead{} &
\colhead{} & \colhead{}
}
\startdata
\multicolumn{8}{l}{\emph{Fitted}} \\
\hline
$M_\mathrm{*}$ & ${\mathrm{M}_\odot}$ & ${0.996}_{-0.035}^{+0.036}$ & 0.963 & \nodata & \nodata & \nodata & \nodata \\
$\pi$ & mas & ${62.486539}_{-0.000066}^{+0.000030}$ & 62.487 & \nodata & \nodata & \nodata & \nodata \\
$\sigma_\mathrm{NEID_{Run~0.5}}$ & ${\mathrm{m~s^{-1}}}$ & ${0.86}_{-0.32}^{+0.45}$ & 0.74 & \nodata & \nodata & \nodata & \nodata \\
$\sigma_\mathrm{NEID_{Run~1}}$ & ${\mathrm{m~s^{-1}}}$ & ${1.11}_{-0.19}^{+0.23}$ & 1.13 & \nodata & \nodata & \nodata & \nodata \\
$\sigma_\mathrm{NEID_{Run~2}}$ & ${\mathrm{m~s^{-1}}}$ & ${1.35}_{-0.13}^{+0.14}$ & 1.35 & \nodata & \nodata & \nodata & \nodata \\
$\sigma_\mathrm{HIRES_{pre}}$ & ${\mathrm{m~s^{-1}}}$ & ${2.85}_{-0.25}^{+0.28}$ & 2.98 & \nodata & \nodata & \nodata & \nodata \\
$\sigma_\mathrm{HIRES_{post}}$ & ${\mathrm{m~s^{-1}}}$ & ${2.56}_{-0.14}^{+0.15}$ & 2.47 & \nodata & \nodata & \nodata & \nodata \\
$\sigma_\mathrm{APF}$ & ${\mathrm{m~s^{-1}}}$ & ${2.12}_{-0.12}^{+0.12}$ & 2.15 & \nodata & \nodata & \nodata & \nodata \\
$\sigma_\mathrm{Hamilton}$ & ${\mathrm{m~s^{-1}}}$ & ${6.58}_{-0.52}^{+0.58}$ & 7.06 & \nodata & \nodata & \nodata & \nodata \\
$\sigma_\mathrm{ELODIE}$ & ${\mathrm{m~s^{-1}}}$ & ${0.077}_{-0.077}^{+4.100}$ & 3.98 & \nodata & \nodata & \nodata & \nodata \\
$\sigma_\mathrm{AFOE}$ & ${\mathrm{m~s^{-1}}}$ & ${2.0}_{-2.0}^{+6.8}$ & 11.5 & \nodata & \nodata & \nodata & \nodata \\
RVZP$_\mathrm{NEID_{Run~0.5}}$ & ${\mathrm{m~s^{-1}}}$ & ${45307.19}_{-0.38}^{+0.38}$ & 45307.17 & \nodata & \nodata & \nodata & \nodata \\
RVZP$_\mathrm{NEID_{Run~1}}$ & ${\mathrm{m~s^{-1}}}$ & ${45303.35}_{-0.40}^{+0.41}$ & 45303.47 & \nodata & \nodata & \nodata & \nodata \\
RVZP$_\mathrm{NEID_{Run~2}}$ & ${\mathrm{m~s^{-1}}}$ & ${45302.95}_{-0.12}^{+0.12}$ & 45302.94 & \nodata & \nodata & \nodata & \nodata \\
RVZP$_\mathrm{HIRES_{pre}}$ & ${\mathrm{m~s^{-1}}}$ & ${1.34}_{-0.11}^{+0.12}$ & 1.33 & \nodata & \nodata & \nodata & \nodata \\
RVZP$_\mathrm{HIRES_{post}}$ & ${\mathrm{m~s^{-1}}}$ & ${2.568}_{-0.058}^{+0.058}$ & 2.54 & \nodata & \nodata & \nodata & \nodata \\
RVZP$_\mathrm{APF}$ & ${\mathrm{m~s^{-1}}}$ & ${-0.956}_{-0.066}^{+0.065}$ & -0.90 & \nodata & \nodata & \nodata & \nodata \\
RVZP$_\mathrm{Hamilton}$ & ${\mathrm{m~s^{-1}}}$ & ${3.86}_{-0.12}^{+0.12}$ & 3.89 & \nodata & \nodata & \nodata & \nodata \\
RVZP$_\mathrm{ELODIE}$ & ${\mathrm{m~s^{-1}}}$ & ${45347.615}_{-0.063}^{+0.066}$ & 45347.690 & \nodata & \nodata & \nodata & \nodata \\
RVZP$_\mathrm{AFOE}$ & ${\mathrm{m~s^{-1}}}$ & ${45350.11}_{-0.48}^{+0.51}$ & 45349.66  & \nodata & \nodata & \nodata & \nodata \\
\hline
\colhead{} & \colhead{} &
\multicolumn{2}{c}{\textbf{HD 190360 b}} &
\multicolumn{2}{c}{\textbf{HD 190360 c}} &
\multicolumn{2}{c}{\textbf{HD 190360 d}} \\
\colhead{} & \colhead{} &
\colhead{Post.} & \colhead{MAP} &
\colhead{Post.} & \colhead{MAP} &
\colhead{Post.} & \colhead{MAP} \\
\hline
\multicolumn{8}{l}{\emph{Fitted}} \\
\hline
$m$ & $\mathrm{M_J}$  & ${1.69}_{-0.15}^{+0.25}$ & 1.54 & ${0.083}_{-0.013}^{+0.085}$ & -- & ${0.0375}_{-0.0054}^{+0.0200}$ & -- \\
$a$ & au  & ${3.965}_{-0.047}^{+0.047}$ & 3.920 & ${0.1298}_{-0.0016}^{+0.0015}$ & 0.1284 & ${0.3886}_{-0.0046}^{+0.0046}$ & 0.3844 \\
$\sqrt{e}\sin\omega$ & --  & ${0.150}_{-0.012}^{+0.011}$ & 0.155 & ${-0.256}_{-0.041}^{+0.046}$ & -0.277 & ${0.04}_{-0.20}^{+0.17}$ & -0.0416 \\
$\sqrt{e}\cos\omega$ & --  & ${0.5582}_{-0.0063}^{+0.0064}$ & 0.555 & ${0.310}_{-0.042}^{+0.039}$ & 0.302 & ${-0.04}_{-0.20}^{+0.21}$ & -0.0557 \\
$i$ & $^\circ$  & ${71}_{-17}^{+42}$ & 72.87 & ${95}_{-51}^{+46}$ & -- & ${88}_{-41}^{+43}$ & -- \\
$\Omega$ & $^\circ$  & ${108}_{-22}^{+60}$ & 108.62 & ${178}_{-119}^{+125}$ & -- & ${177}_{-121}^{+125}$ & -- \\
$\lambda$ & $^\circ$  & ${222.14}_{-0.42}^{+0.41}$ & 222.06 & ${53.7}_{-1.7}^{+1.7}$ & 54.01 & ${50.3}_{-7.9}^{+7.7}$ & 49.15 \\
\\
\multicolumn{8}{l}{\emph{Derived}} \\
\hline
$P$ & yr  & ${7.906}_{-0.010}^{+0.010}$ & 7.904 & ${0.0468627}_{-0.0000011}^{+0.0000011}$ & 0.04686 & ${0.24282}_{-0.00014}^{+0.00014}$ & 0.24285 \\
$P$ & d  & ${2887.5}_{-3.6}^{+3.5}$ & 2886.89 & ${17.11625}_{-0.00039}^{+0.00039}$ & 17.1165 & ${88.690}_{-0.049}^{+0.051}$ & 88.698 \\
$m$ & $\mathrm{M_\oplus}$  & ${537}_{-46}^{+79}$ & 491 & ${26.2}_{-4.0}^{+27.0}$ & -- & ${11.9}_{-1.7}^{+6.2}$ & -- \\
$m\sin i$ & $\mathrm{M_J}$  & ${1.513}_{-0.038}^{+0.038}$ & 1.48 & ${0.0684}_{-0.0021}^{+0.0021}$ & 0.0662 & ${0.0322}_{-0.0025}^{+0.0025}$ & 0.0330 \\
$m\sin i$ & $\mathrm{M_\oplus}$  & ${481}_{-12}^{+12}$ & 468.88 &  ${21.75}_{-0.66}^{+0.67}$ & 21.03 & ${10.23}_{-0.80}^{+0.81}$ & 10.49 \\
$K$ &  ${\mathrm{m~s^{-1}}}$ & ${22.95}_{-0.20}^{+0.20}$ & 22.87 & ${5.49}_{-0.11}^{+0.11}$ & 5.43 & ${1.48}_{-0.11}^{+0.11}$ & 1.54 \\
$\omega$ & $^\circ$  & ${15.0}_{-1.2}^{+1.2}$ & 15.61 & ${320.5}_{-7.7}^{+7.9}$ & 317.49 & ${160}_{-100}^{+115}$ & 216.74 \\
$e$ & --  & ${0.3342}_{-0.0062}^{+0.0061}$ & 0.332 & ${0.163}_{-0.019}^{+0.020}$ & 0.168 & ${0.058}_{-0.040}^{+0.062}$ & 0.00483 \\
$a$ & mas  & ${247.7}_{-2.9}^{+3.0}$ & 244.96 & ${8.11}_{-0.10}^{+0.10}$ & 8.02 & ${24.28}_{-0.29}^{+0.29}$ & 24.02 \\
$T_\mathrm{p}$ & JD  & ${2456423.6}_{-8.9}^{+8.6}$ & 2456428.83 & ${2455210.18}_{-0.35}^{+0.36}$ & 2455210.03 & ${2455234}_{-23}^{+38}$ & 2455238.79 \\
$q$ & --  & ${0.00162}_{-0.00014}^{+0.00024}$ & 0.00153 & ${0.000079}_{-0.000012}^{+0.000082}$ & 0.0000669 & ${0.0000360}_{-0.0000051}^{+0.0000190}$ & 0.0000332 \\
\enddata
\tablecomments{MAP values are not provided for $m_\mathrm{c}$, $m_\mathrm{d}$, $i_\mathrm{c}$, $i_\mathrm{d}$, $\Omega_\mathrm{c}$, nor $\Omega_\mathrm{d}$, as the data do not meaningfully constrain the orbital orientations and true masses of the inner planets. Here, $\pi$ is the stellar parallax, $\sigma$ values are associated RV jitter terms, RVZP values are zero-point offsets, $\Omega$ is the longitude of the ascending node, $\lambda$ is the mean longitude at a reference epoch of 2455197.5 JD, $T_\mathrm{p}$ is the time of periastron passage at the same epoch, and $q$ is the mass ratio.}
\end{deluxetable*}

To establish an interpretation for the $89~\mathrm{d}$ signal without including NEID, we execute a second MCMC analysis with the same parameters as those described in Section \ref{sec:analysis} using both absolute astrometry and RVs. Here, we limit our RV dataset to exclude those from NEID. This ``No NEID" three-planet fit reveals a bimodality in the best-fit period distribution for HD 190360~d, with each peak corresponding to the value reported in \cite{Hirsch_AO} and \cite{2021ApJS..255....8R}, respectively.

\begin{figure}[h!]
    \centering
    \includegraphics[width=\linewidth]{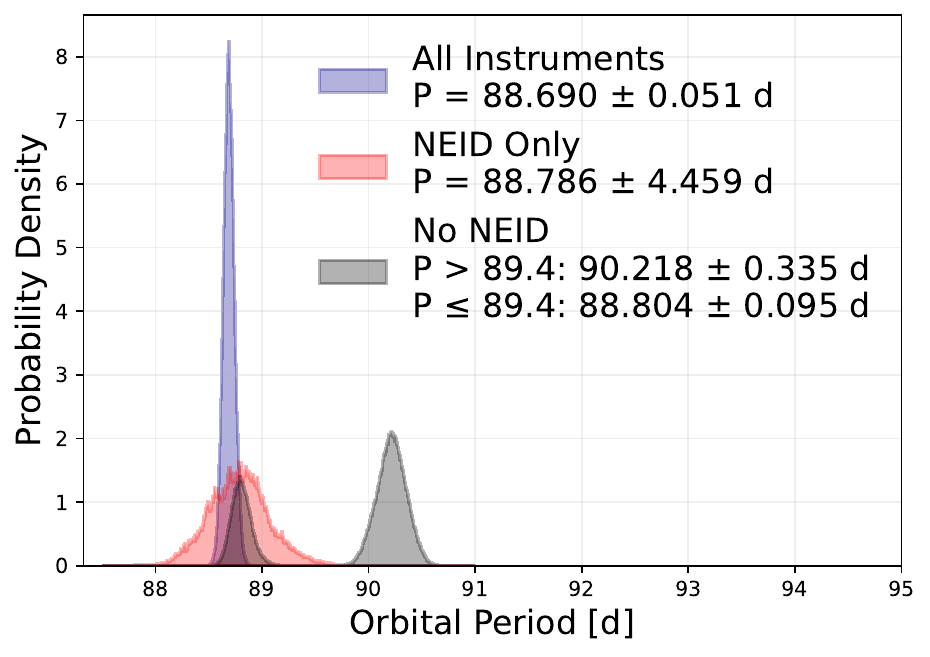}
    \caption{HD 190360~d period distribution for three unique MCMC fits with varying RV datasets. The blue histogram is the period determined from the joint astrometric and RV analysis outlined in Section \ref{sec:analysis} and adopted in this work. Both the red and gray histograms are constructed within the same framework; the red histogram includes only RVs from NEID, whereas the gray histogram includes all RVs except those from NEID. Each case jointly incorporates absolute astrometry from Hipparcos and Gaia. The ``No NEID" fit yields two period peaks at 88.8~d and 90.2~d, but we find that the precision and phase coverage from NEID is sufficient for breaking the bimodality.}
    \label{fig:per_per_inst}
\end{figure}

We lastly consider the capabilities of the NEID RVs to identify the $89~\mathrm{d}$ period induced by HD 190360~d. To do this, we use only the 93 NEID RVs, as well as the absolute astrometry from the HGCA. We again fit for three planets with broad walker initializations and find that our ``NEID Only" fit independently constrains the orbital period for HD 190360~d. In Figure \ref{fig:per_per_inst}, we show these three period distributions, demonstrating NEID's ability to break the bimodality. The bimodality likely arises from the combination of HD 190360~d's low RV semi-amplitude ($K=1.48\pm0.11~\mathrm{m~s^{-1}}$) compared to the $>2~\mathrm{m~s^{-1}}$ archival RV jitter and large observational gaps over 25 years. NEID's $\sim1~\mathrm{m~s^{-1}}$ precision and dense sampling break the period degeneracy.

\begin{figure}[h!]
    \centering
    \includegraphics[width=\linewidth]{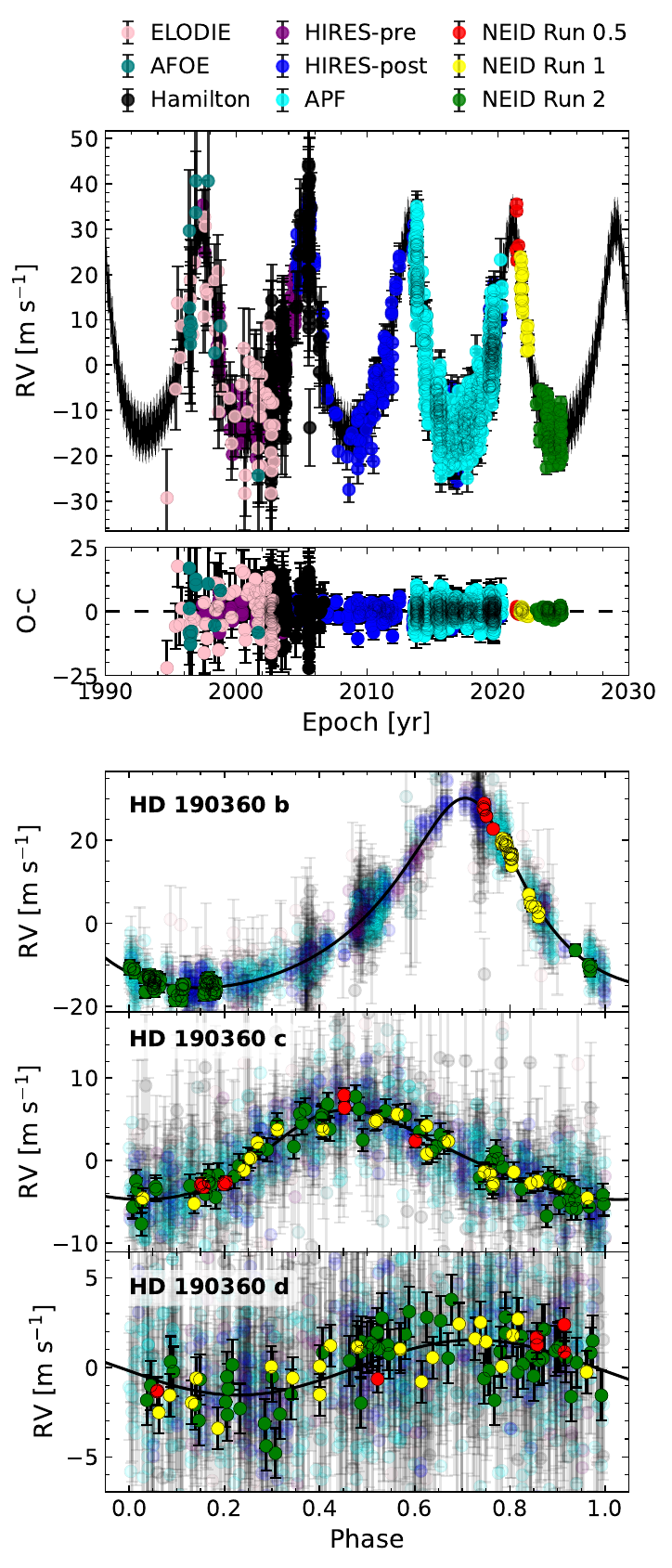}
    \caption{Top panel: RV timeseries for HD 190360. Bottom panel: phase-folded RV diagrams for HD 190360's three known planets. Here, only the NEID data are left opaque to showcase their constraint on each of the planets.}
    \label{fig:RV_phased}
\end{figure}

\begin{figure*}
    \centering
    \includegraphics[width=\linewidth]{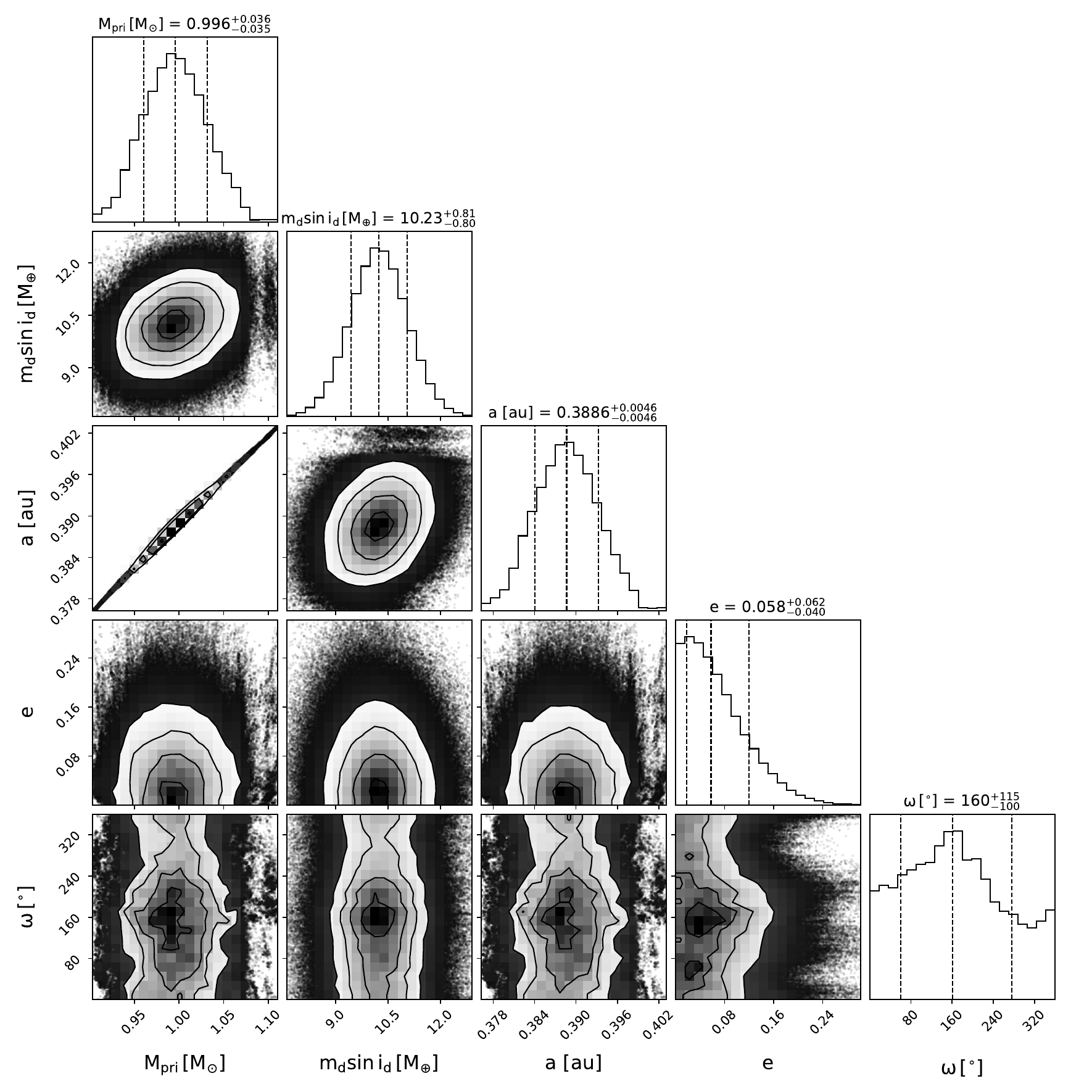}
    \caption{Corner plot of HD 190360~d. From left to right, parameters are stellar mass, minimum planet mass, semimajor axis, eccentricity, and argument of periastron.}
    \label{fig:corner}
\end{figure*}

The phase-folded RV diagram with the NEID RVs highlighted particularly illuminates the presence of HD 190360~d. This, along with the full RV timeseries, is provided in Figure \ref{fig:RV_phased}. A select set of posteriors for HD 190360~d is then shown in Figure \ref{fig:corner}.

Given that the posterior distribution for HD 190360~d's eccentricity has significant probability density near zero ($e_\mathrm{d}={0.058}_{-0.040}^{+0.062}$) and is weakly constrained ($e_{\mathrm{d}} < 0.256$ at $3\sigma$), we also considered a model in which $e_\mathrm{d}=0$. Following the same fitting procedure described in Section \ref{sec:analysis}, this constraint produces nearly identical posterior distributions for the parameters of all three planets. The maximum log-likelihood of the circular solution for HD 190360~d ($\log\hat{\mathcal{L}}=-2457.80$) is statistically indistinguishable from that of the free-eccentricity model ($\log\hat{\mathcal{L}}=-2457.98$). The circular model yields a modestly lower BIC owing to its reduced number of free parameters, but allowing HD 190360~d's eccentricity to vary provides a more flexible and physically motivated model that ensures that uncertainties in orbital parameters are propagated consistently. Throughout this work, we report parameters from the model in which $e_\mathrm{d}$ is a free parameter.

The mass-inclination degeneracy for HD 190360~b is broken by including the astrometry from Hipparcos and Gaia. As reported in Table \ref{tab:planet_posteriors}, HD 190360~b has a returned inclination of $i_\mathrm{b}={71}_{-17}^{+42\circ}$ and a maximum-likelihood inclination of $i_\mathrm{b}=72.87^\circ$. While the astrometric constraints from Hipparcos and Gaia are not sufficient for revealing the inclinations of either HD 190360~c or HD 190360~d, the constraint on the outer planet naturally motivates testing whether a model that explicitly imposes coplanarity is consistent with the data.

We therefore performed a fit in which all three planets were forced to share a common orbital plane ($i_\mathrm{d}=i_\mathrm{c}=i_\mathrm{b}$ and $\Omega_\mathrm{d}=\Omega_\mathrm{c}=\Omega_\mathrm{b}$), leaving $i_\mathrm{b}$ and $\Omega_\mathrm{b}$ as free parameters for the data to constrain. The resulting maximum log-likelihood of the coplanar solution ($\log\hat{\mathcal{L}}=-2457.45$) is consistent with that of the nominal model ($\log\hat{\mathcal{L}}=-2457.98$). In this coplanar framework, we find true masses of $m_\mathrm{c}={24.5}_{-2.4}^{+4.1}~\mathrm{M_\oplus}$ and $m_\mathrm{d}={11.6}_{-1.4}^{+2.0}~\mathrm{M_\oplus}$. Since the data cannot distinguish between the two orbital configurations, and in the absence of direct observational evidence linking the orbital planes of HD 190360~b and the two inner planets, we retain the more general model presented in Table \ref{tab:planet_posteriors}, in which the orbital planes are allowed to vary independently.




\section{Discussion} \label{sec:discussion}

\subsection{89~d Period vs. Known Activity Cycles}

\begin{figure}[h!]
    \centering
    \includegraphics[width=\linewidth]{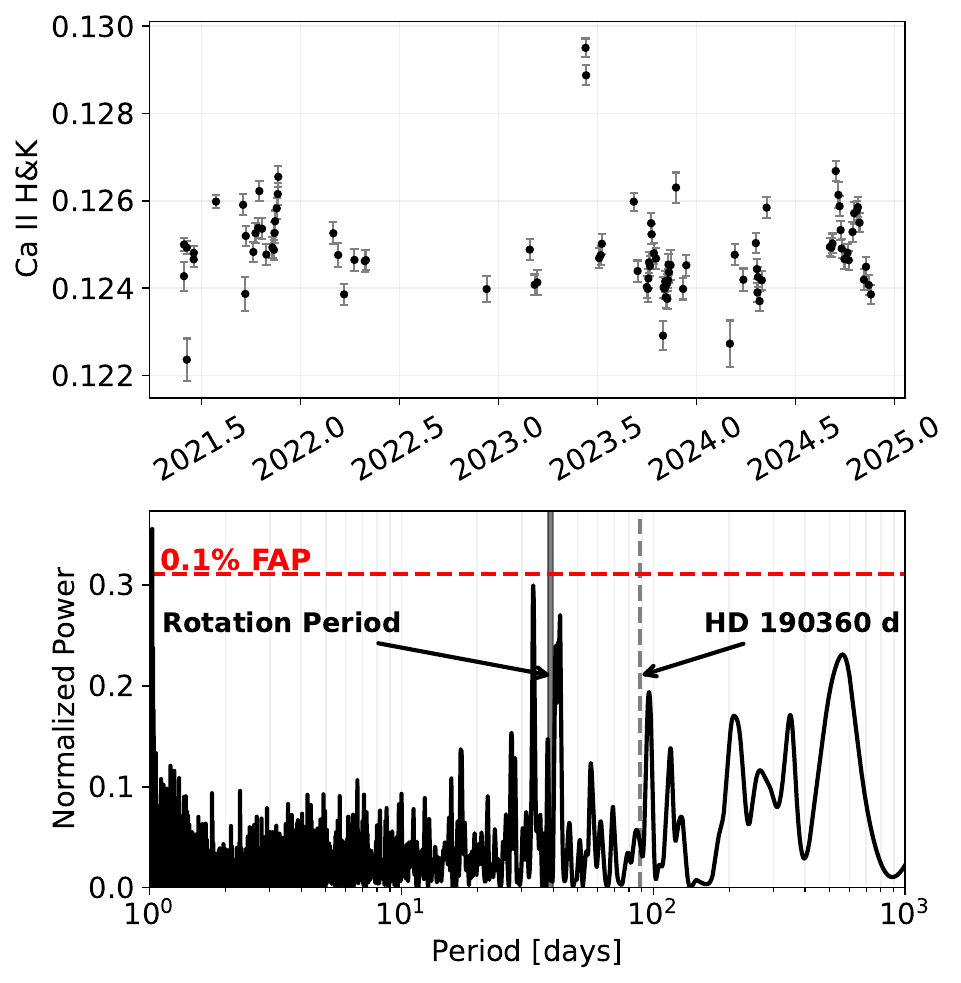}
    \caption{Top: timeseries of NEID Ca II H and K values. Bottom: Lomb-Scargle periodogram for the same dataset. The horizontal dashed line marks the 0.1\% false alarm probability (FAP) level, computed using \texttt{astropy}'s Lomb-Scargle routine following the approach outlined in \cite{2008MNRAS.385.1279B}. We identify no strong periods at or near the signal reported for HD 190360~d.}
    \label{fig:activities}
\end{figure}

Periodic RV signals can be produced by planets, stellar activity, or systematics, with the potential for each to mimic the others and obscure the true origin of the observed variability \citep[e.g.,][]{2008ApJ...683L..63W, 2024AJ....167..243B}. Recovering the source of such RV signals is particularly challenging in the case of a small amplitude, like HD 190360~d. Previously, \cite{Baum_2022} and \cite{Obridko_2022} used Ca II H and K activity indicators but observed no significant variation for HD 190360. Here, we examine all available activity indicators obtained from our NEID spectra derived using the NEID DRP (including Ca II H and K, NaI, and H$\alpha$), as well as Ca II H and K lines from HIRES spectra published in \cite{2021ApJS..255....8R}. For each indicator, we assess periodic variability using Lomb-scargle periodograms and find no significant power at or near the 89~d period determined for HD 190360~d. Figure \ref{fig:activities} highlights NEID's observed Ca II H and K values and subsequent periodogram, where the stellar rotation period (see Section \ref{sec:star}) may be evident. Lastly, we find no evidence of variability in the TESS photometry. Given that our best-fit period of $P_\mathrm{d}=88.686^{+0.053}_{-0.050}~\mathrm{d}$ cannot be explained by a known stellar signal, and that it has independently been determined across multiple instrument datasets, we conclude HD 190360~d to be a \textit{bona fide} planet.

\subsection{Long-term Stability}

\begin{figure}
    \centering
    \includegraphics[width=\linewidth]{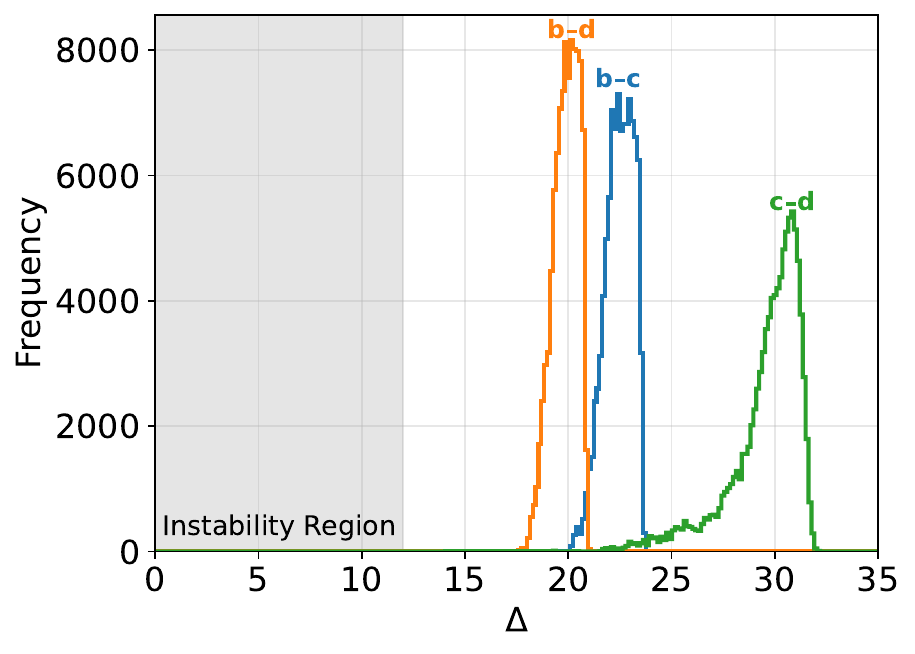}
    \caption{Dynamical separation distributions for the three combinations of interacting planets. We observe that all three combinations have $\Delta>12$ by a relatively wide margin. A priori, interactions between HD 190360~c and HD 190360~d would be expected to be the largest, and so their $\Delta$ distribution being the highest is a promising indicator that this configuration of planets would have been long-term stable.}
    \label{fig:hill}
\end{figure}

Dynamical stability is a key consideration in multi-planet systems, particularly around field-age stars that have evolved the longest since formation. To estimate the likelihood of long-term stability in the HD 190360 planetary system, we compute the dynamical separations ($\Delta$) between our three planets in units of their mutual Hill radii following Eqn. 1 from \cite{2024AJ....167..271V} using the posterior values from Table \ref{tab:planet_posteriors}. Planets with $\Delta>12$ are generally more likely to remain stable on Gyr timescales, although instabilities can still occur at larger separations depending on the specific system architecture \citep[e.g.,][]{1996Icar..119..261C, 2015ApJ...807...44P, 2017Icar..293...52O}. To estimate distributions of our planets' dynamical separations, we sample the 100,000 most likely orbits from the fit described in Section \ref{sec:results}. We note that long-term stability criteria for multi-planet systems with high eccentricities or varying inclinations are not well understood analytically or numerically \citep[e.g.,][]{2018AJ....156...95H, 2024ApJ...973..108B}. However, given that the system is far from the stability boundary (see Figure \ref{fig:hill}) and does not have extreme eccentricities or inclinations, we ignore secular evolution and deem the HD 190360 planetary system to be long-term stable. Additionally, and according to the maximum-likelihood values provided in Table \ref{tab:planet_posteriors}, HD 190360~c's apastron (0.150~au) and HD 190360~d's periastron (0.383~au) are widely-separated enough that we would not expect strong gravitational interactions. Similarly, HD 190360~d's apastron (0.386~au) and HD 190360~b's periastron (2.618~au) are such that mutual interaction is not expected.

To complement this analytic stability assessment, we employed the Stability of Planetary Orbital Configurations Klassifier (SPOCK; \citealt{2020PNAS..11718194T}), a machine-learning classifier that uses the gradient-boosted decision tree algorithm \texttt{XGBoost} \citep{2016arXiv160302754C} trained on $N$-body integrations to predict the long-term stability of multi-planet systems. We evaluated the 10,000 highest-likelihood posterior samples from our MCMC chains and find that 92.4\% have predicted stability probabilities $p_\mathrm{stable}>0.5$, with a median of $p_\mathrm{stable}=0.90$. SPOCK predicts stability over $10^9$ orbits of the system's innermost planet ($\sim50~\mathrm{Myr}$ for HD 190360). We find that the primary predictor of instability in our posterior samples is the eccentricity of HD 190360~d, with unstable configurations predominantly occurring at $e_\mathrm{d}>0.08$. These results corroborate our Hill-radius-based assessment and suggest that our orbital solution for the HD 190360 planetary system is long-term dynamically stable.

HD 190360 was identified as a promising candidate for future observations with the Habitable Worlds Observatory by \citep{HWO_targlist}. Assuming the only two planets known at the time, \cite{2024AJ....168..195K} injected a terrestrial-mass planet into the habitable zone of HD 190360 and determined a region of orbital distances that are dynamically viable. Here, we inject a hypothetical planet e with $1~\mathrm{M_\oplus}$ at $1~\mathrm{au}$ and apply the stability criteria outlined above. For this hypothetical planet, we find dynamical separations of $\Delta_\mathrm{be}=14.9$, $\Delta_\mathrm{ce}=54.1$, and $\Delta_\mathrm{de}=38.7$, suggesting that an Earth-mass planet with Earth-like separation from HD 190360 could be long-term stable.

\subsection{HD 190360~d: Super-Earth or Sub-Neptune?} \label{subsec:nomenclature}


The precise nomenclature of exoplanet classification is an important foundation for our interpretation of planet populations \citep[see, for example,][]{2018ApJ...856..122K, 2024arXiv240909666P}. Often inherited from our solar system's planets, these labels are extended to exoplanets based on limited observables and sometimes obscure important differences in composition or atmospheric properties. The boundary between super-Earths and sub-Neptunes, often associated with the radius valley \citep{2017AJ....154..109F}, remains particularly difficult to distinguish.

Super-Earths are considered to be rocky planets with small atmospheres and $R\lesssim1.75~\mathrm{R_\oplus}$ (\citealt{2007ApJ...669.1279S}, \citealt{2008ApJ...673.1160A}, \citealt{2015ApJ...801...41R}, \citealt{2017AJ....154..109F}), and sub-Neptunes to be larger planets ($R\gtrsim1.75~\mathrm{R_\oplus}$) whose atmospheres contribute more significantly to their total mass. Using mass-radius relationships from \cite{mass_rad1}, \cite{mass_rad2} and references therein, a mass of $10~\mathrm{M_\oplus}$ yields an estimated radius for HD 190360~d between $2.0-3.5~\mathrm{R_\oplus}$. While this physical size suggests a more Neptune-like planet, HD 190360~d's true classification is ambiguous without an empirical estimate of its radius and true mass measurement. Since the astrometric amplitude of HD 190360~d is $\lesssim1~\mu\mathrm{as}$, significantly below the anticipated sensitivity of Gaia DR4 \citep[e.g.,][]{2024A&A...686L...2G}, a direct astrometric mass measurement is unlikely in the near future. A growing consensus suggests that planets with $m\gtrsim10~\mathrm{M_\oplus}$ may not be able to sustain properties that would lead them to being categorized as terrestrial \citep[e.g.,][]{2010ApJ...722.1854W, 2024A&A...688A..59P}. We also note that mass-radius relations are not a universally predictive tool, and that planets with sub-Neptune-like masses ($m>10~\mathrm{M_\oplus}$) and super-Earth-like radii ($R<1.75~\mathrm{R_\oplus}$) have been found \citep[e.g.,][]{2014ApJS..210...25X}. Assuming a mutual inclination with HD 190360~b ($i_\mathrm{b}={71}_{-17}^{+42\circ}$), HD 190360~d would have a true mass of $m_\mathrm{d}={11.5}_{-1.3}^{+1.9}~\mathrm{M_\oplus}$. By adopting the measured $m\sin i$ from Table \ref{tab:planet_posteriors} and assuming inclinations drawn randomly from a $\cos i$ distribution, we estimate an $\approx82\%$ probability that the true mass of HD 190360~d is less than that of Neptune.



\subsection{Comparison to Other Multi-planet Systems}

\begin{figure}
    \centering
    \includegraphics[width=\linewidth]{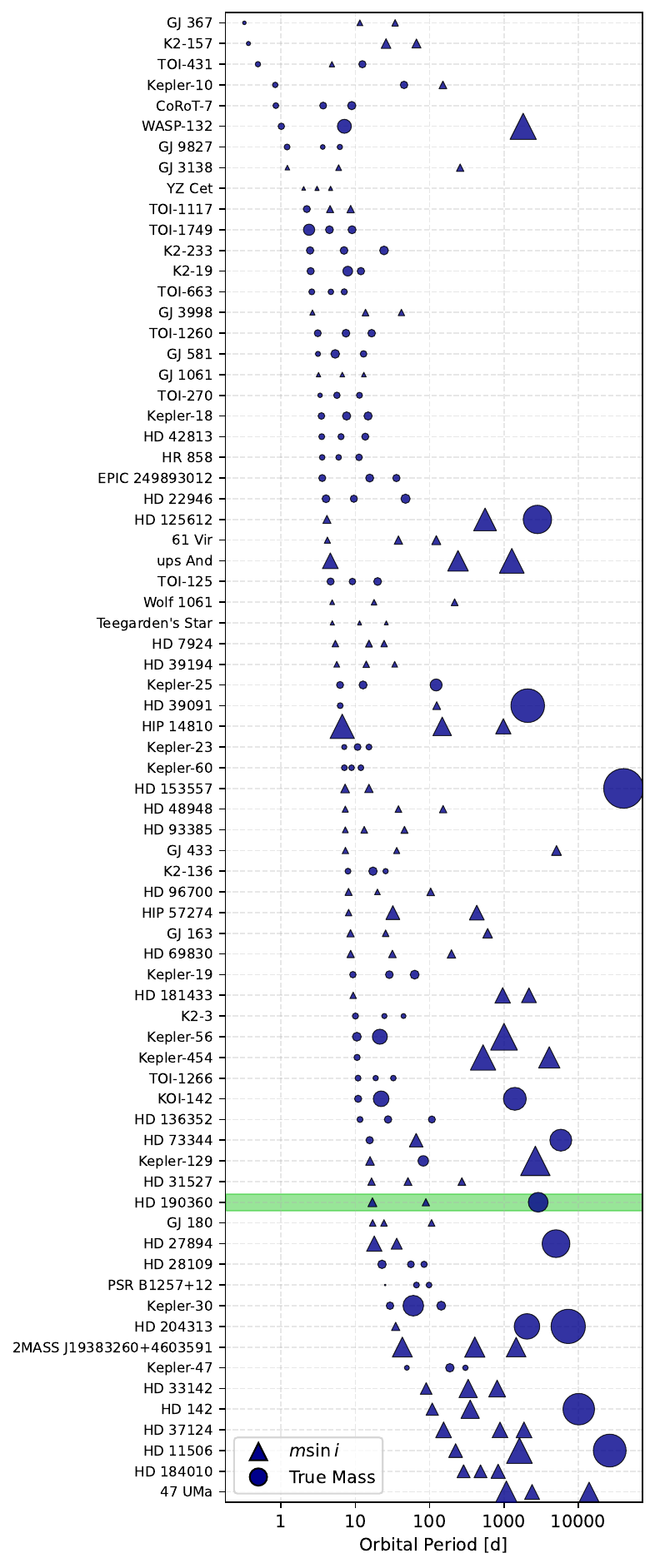}
    \caption{Orbital architectures of known three-planet systems in the NASA Exoplanet Archive. Host stars are shown along the y-axis, with planet orbital periods on the x-axis. Marker size scales with planet mass, and marker shape distinguishes $m\sin i$ measurements (triangles) from true masses (circles). The HD~190360 system is highlighted in green.}
    \label{fig:3p_panel}
\end{figure}

For the first time, the architecture of the HD 190360 planetary system can be discussed in the context of three-planet systems. We identify 71 other known three-planet systems in the NASA Exoplanet Archive \citep{2025PSJ.....6..186C} where all companions have published masses and periods. We present these systems in Figure \ref{fig:3p_panel} and summarize their mass configurations in Table \ref{tab:mass_configurations}. The prevalence of configurations in which the least massive planet occupies the shortest-period orbit may reflect a trend in multi-planet systems where outer giants facilitate the formation of smaller, inner planets \citep[e.g.,][]{bryan2024friendsfoesstrongcorrelation, zhu2024metallicitydimensionsuperearthcold}. Conversely, the rarity of configurations where the least massive planet resides in the longest-period orbit may point to dynamical instabilities arising from disk-driven migration \citep[e.g.,][]{2002A&A...394..241T}. Of course, this may also be related to our insensitivity to low-mass planets at longer periods.

\begin{table}[ht]
\centering
\caption{Mass configurations in known three-planet systems}
\begin{tabular}{ccc}
\hline
\textbf{Configuration} & \textbf{Count} & \textbf{\% of 3-planet systems} \\
\hline
(3, 2, 1) & 26 & 36.6\% \\
(3, 1, 2) & 21 & 29.6\% \\
\textbf{(2, 3, 1)} & \textbf{14} & \textbf{19.7\%} \\
(1, 3, 2) & 5  & 7.0\% \\
(1, 2, 3) & 4  & 5.6\% \\
(2, 1, 3) & 2  & 2.8\% \\
\hline
\end{tabular}
\tablecomments{Each configuration tuple indicates the ordering of three planets from shortest to longest orbital period, with the numbers 1 to 3 denoting the most to least massive planet in the system. The second column gives the number of systems exhibiting each configuration, while the rightmost column shows the corresponding percentage relative to all 71 three-planet systems considered here. The (2, 3, 1) configuration is made bold to highlight it as the configuration for HD 190360.}
\label{tab:mass_configurations}
\end{table}

The observed architecture of the HD 190360 planetary system serves as a fossil of its formation and evolutionary history. The outer planet's semimajor axis ($a_\mathrm{b}=3.965\pm0.047~\mathrm{au}$) places it beyond the range for ``lukewarm" giants ($a=0.3-3~\mathrm{au}$, $m>120~\mathrm{M_\oplus}$), a region \cite{2022ApJS..262....1R} showed may suppress inner planet formation. The presence of both HD 190360~c and HD 190360~d suggests that HD 190360~b formed beyond this zone. Moreover, HD 190360~b's moderate eccentricity ($e_\mathrm{b}=0.3342^{+0.0061}_{-0.0062}$) may reflect past dynamical interactions or secular perturbations, while the lower eccentricities of the inner planets ($e_\mathrm{c}=0.163^{+0.02}_{-0.019}$, $e_\mathrm{d}=0.058^{+0.062}_{-0.040}$) suggest they avoided significant perturbations during this excitation period. \cite{2022ApJS..262....1R} also found that outer companions to smaller inner planets typically exhibit eccentricities $\lesssim0.4$, with which the architecture of the HD 190360 system is indeed consistent. HD 190360 joins a growing number of similarly sculpted multi-planet systems with eccentric outer giants and small inner companions \citep[e.g.,][]{2017A&A...602L...8T, 2024ApJS..270....8W, 2025AJ....169..200Z} that support the theory that cold giants can harbor small inner planets.

\begin{figure*}
    \centering
    \includegraphics[width=\linewidth]{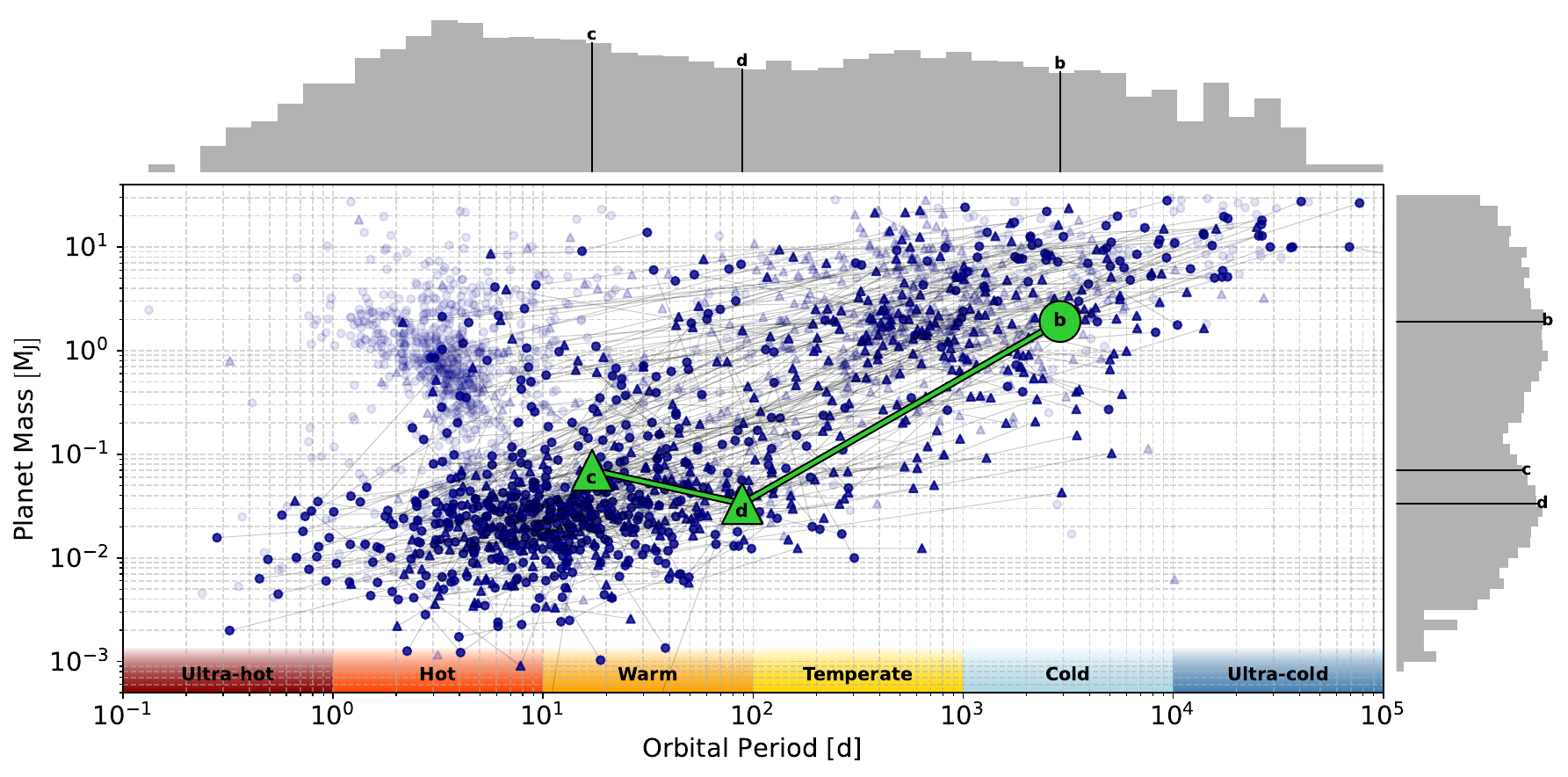}
    \caption{All exoplanets from the NASA Exoplanet Archive with known mass and orbital period. Triangle and circle point markers represent minimum and true masses, respectively. The larger green markers correspond to HD 190360's three planets. All single-planet systems are made transparent, while all multi-planet systems are connected with a thin black line in order from shortest to longest period. Named period-bins are provided for reference according to rough temperature scales for planets orbiting main sequence FGK stars, and correspond to $0<P~\mathrm{\left[d\right]}\leq1$ (ultra-hot), $1<P~\mathrm{\left[d\right]}\leq10$ (hot), $10<P~\mathrm{\left[d\right]}\leq100$ (warm), $100<P~\mathrm{\left[d\right]}\leq1000$ (temperate), $1000<P~\mathrm{\left[d\right]}\leq10^4$ (cold), and $10^4<P~\mathrm{\left[d\right]}\leq10^5$ (ultra-cold). The gray histograms at the top and right correspond to the observed distributions of orbital periods and masses.}
    \label{fig:mass_per}
\end{figure*}

The HD 190360 planetary system comprises three planets that span a range of both masses and orbital periods (see Figure \ref{fig:mass_per}). With a true dynamical mass and well-known orbit, HD 190360~b is a benchmark giant planet firmly within the landscape of cold Jupiters. HD 190360~c is a short-period planet with $m_\mathrm{c}\sin i_\mathrm{c}={21.75}_{-0.66}^{+0.67}~\mathrm{M_\oplus}$, and HD 190360~d is found to have a minimum mass that lies within one of the most common ranges observed among exoplanets, though its orbital period places it within a local period valley \citep{per_valley1, per_valley2}. The dearth of planets in this period range is suspected to be physical for giant planets, but may be a result of an observational bias for lower-mass planets \citep{2010ApJ...722.1854W}.

\section{Conclusion} \label{sec:conclusion}

We have presented the detection and confirmation of HD 190360~d, a low-mass ($m_\mathrm{d}\sin i_\mathrm{d}=10.23^{+0.81}_{-0.80}~\mathrm{M_\oplus}$), warm ($P_\mathrm{d}={88.690}_{-0.049}^{+0.051}~\mathrm{d}$) planet orbiting a Sun-like star in the solar neighborhood ($d=16.0~\mathrm{pc}$). HD 190360~d is among the 50 lowest RV semi-amplitude signals confirmed for a non-transiting planet ($K_\mathrm{d}=1.48\pm0.11~\mathrm{m~s^{-1}}$), fewer than 25 of which are FGK-type stars. While low-RV amplitude planets remain a challenge to confirm, HD 190360~d is an exciting testament to the capabilities of current extreme precision spectrographs like NEID. This detection was enabled by 93 high-precision NEID RVs, 54 more than in \cite{NETSII}, which most recently identified the signal as a planet candidate. The additional observations were critical for breaking the period degeneracy present in archival RV data (see Figure \ref{fig:per_per_inst}). Bayesian model comparison strongly favors the three-planet model ($\Delta\mathrm{BIC}=105.65$) and eliminates the peak in the residual periodogram presented in \cite{NETSII}. We verify this signal is not caused by stellar activity, enabling a comprehensive characterization of the HD 190360 planetary system's architecture.

HD 190360~d contributes to an emerging population of low-mass, short- to intermediate-period planets in systems with cold Jupiters. In agreement with current theories on the formation of our own solar system, it is thought that the presence of a giant outer planet may harbor stability within the inner disk for shorter-period planets to form. Systems with known cold Jupiters may therefore be fruitful laboratories for searching for lower-amplitude planets. The architecture of the HD 190360 system suggests past dynamical interactions that excited HD 190360~b's orbit while preserving the inner system.

The fourth data release of Gaia is anticipated to be released by the end of 2026. One exciting data product from this release is 66 months of the mission's epoch astrometry, which will enable the detection of intra-decade astrometric motion of the host star at micro-arcsecond precision. For HD 190360, this will not only mean a higher-fidelity solution to HD 190360~b's orbit, but may also reveal planetary companions with orbital periods between that of HD 190360~d ($P_\mathrm{d}=88.690^{+0.051}_{-0.049}~\mathrm{d}$) and HD 190360~b ($P_\mathrm{b}=7.906\pm0.010~\mathrm{yr}$). With a dynamically stable habitable zone, HD 190360 will remain a compelling target for studies aimed at discovering temperate planets in the period range of 100-1000~d.

\begin{acknowledgments}

The authors thank the anonymous referee for their thoughtful and constructive review that enhanced the quality of this manuscript. M.R.G. would like to thank Tiger Lu for his careful review of our system's long-term stability arguments. This work has made use of the Amherst College High-Performance Computing System.

NEID data presented herein were obtained at the WIYN Observatory from telescope time allocated to NN-EXPLORE through the scientific partnership of the National Aeronautics and Space Administration, the National Science Foundation, and the National Optical Astronomy Observatory. NEID is funded by NASA through JPL contract 1547612 and the NEID Data Reduction Pipeline is funded through JPL contract 1644767. NN-EXPLORE is managed by the Jet Propulsion Laboratory, California Institute of Technology under contract with the National Aeronautics and Space Administration.  This work was performed for the Jet Propulsion Laboratory, California Institute of Technology, sponsored by the United States Government under the Prime Contract 80NM0018D0004 between Caltech and NASA.

Based in part on observations at Kitt Peak National Observatory, NSF’s NOIRLab (Prop. ID 2021A-2015, 2021B-2015, 2022A-2015, PI: S. Mahadevan; 2021B-0225, 2022A-923895, PI: A. Lin), managed by the Association of Universities for Research in Astronomy (AURA) under a cooperative agreement with the National Science Foundation. The authors are honored to be permitted to conduct astronomical research on Iolkam Du\'ag (Kitt Peak), a mountain with particular significance to the Tohono O\'odham. We thank the NEID Queue Observers and WIYN Observing Associates for their skillful execution of the NEID observations.

This research has made use of the SIMBAD database and the VizieR catalogue access tool, both operated at CDS, Strasbourg, France. Additionally, our work has made use of the NASA Exoplanet Archive, which is operated by the California Institute of Technology, under contract with the National Aeronautics and Space Administration under the Exoplanet Exploration Program. This work has made use of data from the European Space Agency (ESA) mission Gaia (\href{https://www.cosmos.esa.int/gaia}{https://www.cosmos.esa.int/gaia}), processed by the Gaia Data Processing and Analysis Consortium (DPAC, \href{https://www.cosmos.esa.int/web/gaia/dpac/consortium}{https://www.cosmos.esa.int/web/gaia/dpac/consortium}). Funding for the DPAC has been provided by national institutions, in particular the institutions participating in the Gaia Multilateral Agreement.

The Center for Exoplanets and Habitable Worlds is supported by Penn State and its Eberly College of Science.



\end{acknowledgments}

%

\facilities{OHP:1.93m (ELODIE), FLWO:1.5m (AFOE), Keck:I (HIRES), Automated Planet Finder (Levy),
Lick (Hamilton), WIYN (NEID), HIPPARCOS, Gaia}


\software{\texttt{orvara} \citep{orvara}, \texttt{emcee} \citep{emcee}, \texttt{corner} \citep{2016JOSS....1...24F}, \texttt{binary\_mc} \citep{mark_giovinazzi_2025_15352084}, \texttt{astropy} \citep{astropy2013, astropy2018}, \texttt{NumPy} \citep{harris2020array}, \texttt{matplotlib} \citep{Hunter:2007}}




\bibliography{main}{}
\bibliographystyle{aasjournal}



\end{document}